%% file: Conference.tex
\documentclass[conference]{IEEEtran}
\IEEEoverridecommandlockouts

\input{preamble}
\begin{document}
\pagenumbering{gobble}
\title{Flexible RISs: Learning-based Array Manifold Estimation and Phase-shift Optimization}
\author{
\IEEEauthorblockN{Mohamadreza Delbari$^1$, Ehsan Mohammadi$^1$, Mostafa Darabi$^{1,2}$,\\ Arash Asadi$^3$, Alejandro Jiménez-Sáez$^1$, and Vahid Jamali$^1$}\\
\IEEEauthorblockA{
$^1$TU Darmstadt, Germany,
$^2$The University of British Columbia, Canada, $^3$Delft University of Technology, Netherlands\\
\thanks{The work of M. Delbari, and V. Jamali was supported in part by the Deutsche Forschungsgemeinschaft (DFG, German Research Foundation) within the Collaborative Research Center MAKI (SFB 1053, Project-ID 210487104), in part by the LOEWE initiative (Hesse, Germany) within the emergenCITY Centre under Grant LOEWE/1/12/519/03/05.001(0016)/72, and in part by the German Federal Ministry for Research, Technology and Space (BMFTR) under the program of ``Souverän. Digital. Vernetzt.'' joint project Open6GHub plus (Project-ID 16KIS2407). Jim\'{e}nez-S\'{a}ez's work was supported by the Deutsche Forschungsgemeinschaft (DFG, German Research Foundation) – Project-ID 287022738 – TRR 196 MARIE within project C09. Asadi's work was in part supported by DFG HyRIS (455077022) and DFG mmCell (416765679).
}
}
}
\maketitle
\begin{abstract}
\Glspl{RIS} are envisioned as a key enabler for next-generation wireless networks, offering programmable control over propagation environments. While extensive research focuses on planar \gls{RIS} architectures, practical deployments often involve non-planar surfaces, such as structural columns or curved facades, where standard planar beamforming models fail. Moreover, existing analytical solutions for curved \glspl{RIS} are often restricted to specific, pre-defined array manifold geometries. To address this limitation, this paper proposes a novel \gls{DL} framework for optimizing the phase shifts of non-planar \glspl{RIS}. We first introduce a low-dimensional parametric model to capture arbitrary surface curvature effectively. Based on this, we design a \gls{NN} that utilizes a sparse set of received power measurements to estimate the surface geometry and derive the optimal phase configuration. Simulation results demonstrate that the proposed algorithm converges fast and significantly outperforms conventional planar beamforming designs, validating its robustness against arbitrary surface curvature. We also analyze the impact of the measurement location error on the algorithm's performance.
\end{abstract}
\IEEEpeerreviewmaketitle
\glsresetall
\section{Introduction}
\label{Introduction}
\Gls{RIS} is a promising technology for next-generation wireless networks, offering programmable control over wireless propagation channels~\cite{Wu2019,Basar2019,Delbari2026fast}. To achieve this, \glspl{RIS} must be configured accurately. \Glspl{RIS} typically comprise a large number of programmable elements that apply specific phase shifts to incoming signals to reflect them in a desired direction/location. While most existing literature models \glspl{RIS} as planar surfaces, this idealized assumption may not hold in certain practical settings where \glspl{RIS} may need to be mounted on non-planar surfaces, such as structural columns~\cite{Chen2025}, vehicle bodies~\cite{Tagliaferri2022,Mizmizi2023}, or curved building facades~\cite{wang2023flexible, he2022tile}. To address this challenge, \textit{flexible \glspl{RIS}} are proposed and successfully implemented in \cite{wang2023flexible,he2022tile}. However, if an \gls{RIS} is implemented on a non-planar surface but its phase shifts are optimized based on a planar assumption, system performance, e.g., in terms of data rate, sensing accuracy, or tracking capability, significantly degrades due to the geometric mismatch. Under these conditions, conventional theoretical phase shift optimization may be inefficient.

Although some studies have considered curved \glspl{RIS} and optimized phase shifts analytically, they are typically limited to specific structural geometries~\cite{mizmizi2022conformal,Mursia2025}. It is also important to distinguish the scope of this work from \textit{flexible and movable \gls{RIS}} architectures discussed in~\cite{An2025,Ren2025,ranasinghe2025flexible}. While those studies assume that any individual \gls{RIS} element locations are known and controllable in addition to their phase shifts, we consider scenarios where the \gls{RIS} is conformal to a static surface. In our model, element positions are constrained by the underlying geometry and are treated as fixed but unknown variables. This geometry estimation can be formulated as a non-linear, non-convex optimization problem. However, solving this problem iteratively via numerical methods during real-time implementation is computationally prohibitive. \Gls{DL} methods have received significant attention for wireless communication tasks, offering lower real-time computational complexity and the ability to solve problems that are analytically intractable~\cite{Huang2020,ElMossallamy2020}. For example, \cite{Peng2025} optimized \gls{RIS} phase shifts under partial \gls{CSI} availability, and \cite{ramezani2025machine} exploited \gls{DL} for \gls{RIS} \gls{NF} localization. Therefore, in this paper, we develop a \gls{NN}-based method to enable real-time optimization of \gls{RIS} phase shifts on non-planar surfaces with unknown array manifold geometry.

In this work, we train an \gls{NN} to implicitly predict the surface geometry and optimize the phase shifts. The input to our \gls{NN} consists of power measurements taken at various locations when conventional beamforming methods are applied to the surface. To the best of the authors’ knowledge, non-planar \gls{RIS} phase shift optimization using \glspl{NN} has not yet been investigated in the literature. Our key contributions are as follows:
\begin{itemize}
    \item First, we introduce a general parametric model to capture the curvature behavior of the \gls{RIS}, effectively reducing the dimensionality of the problem from the total number of \gls{RIS} elements to a smaller subset of variables.
    \item Next, based on this model, we propose a \gls{DL} method that, by sampling received power at different \gls{MU} locations, first estimates the geometry of the curved \gls{RIS} and subsequently optimizes its phase shifts.
    \item Finally, we evaluate the performance of the proposed method against benchmarks. We discuss the convergence behavior of the employed \gls{NN} and compare its performance with a planar design as the variance of the element locations increases. We also show how much the \gls{NN} is robust against the location error.
\end{itemize}

\begin{figure}
    \centering
    \includegraphics[width=0.3\textwidth]{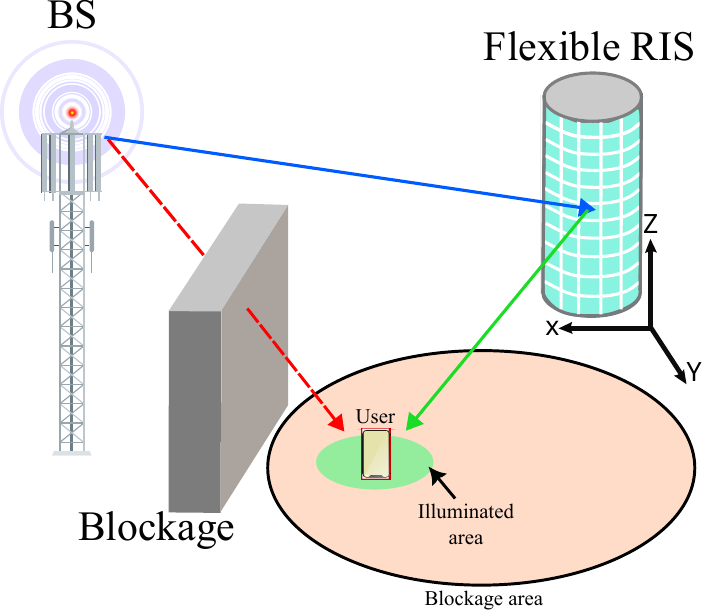}
    \caption{Schematic illustration of the considered RIS-assisted downlink communication system.}
    \label{fig:system model}
    \vspace{-0.5 cm}
\end{figure}

\textit{Notation:} Bold capital and small letters are used to denote matrices and vectors, respectively.  $(\cdot)^\Trans$, $(\cdot)^\Herm$, and $\Uset(a,b)$ denote the Transpose, Hermitian, and Uniform distribution from $a$ to $b$, respectively. Furthermore, $[\bA]_{m,n}$ and $[\ba]_{n}$ denote the element in the $m$th row and $n$th column of matrix $\bA$ and the $n$th entry of vector $\ba$, respectively. $\Rset$ and $\Cset$ represent the sets of real and complex numbers, respectively. Finally, $\jj$ is the imaginary unit, and $\Ex\{\cdot\}$  represents expectation, and $\mathcal{CN}(\bmu,\bSigma)$ denotes a complex Gaussian random vector with mean vector $\bmu$ and covariance matrix $\bSigma$. 


\section{System, Channel and RIS Model}
In this section, we first introduce the considered system model for $K$ \glspl{MU}. Subsequently, we describe the channel model along with the considered \gls{RIS} model.
\subsection{System Model}
\label{System Model}
In this paper, we study a narrow-band downlink communication scenario consisting of a \gls{BS} equipped with $N_t$ transmit antennas, an \gls{RIS} with $N$ unit cells, and $K$ single-antenna \glspl{MU} (or equivalently $K$ different locations of the same \gls{MU}). The $k$th received signal, $y_k\in\Cset$ is given by:
\begin{equation}
\label{Eq: system model}
	y_{k} = \big(\bh_{d,k}^\Herm + \bh_{r,k}^\Herm \bGamma \bH_t \big) \bx +n_k,
\end{equation}
where $\bx\in\Cset^{N_t}$ denotes the transmit signal vector, and $n_k\in\Cset$ is the \gls{AWGN} at the $k$th \gls{MU}, i.e., $n_k\sim\sCN(0,\sigma_n^2)$, where $\sigma_n^2$ is the noise power. Assuming a linear beamforming, we can write the transmit vector $\bx=\bq s$, where $\bq\in\Cset^{N_t}$ is the beamforming vector and $s\in\Cset$ is the data symbol. Here, $\Ex\{|s|^2\}=1$ holds so $\|\bq\|^2\leq P_t$ must hold to satisfy the transmit power constraint as $P_t$ denotes the maximum transmit power. Additionally, $\bh_{d,k} \in \Cset^{N_t}$,  $\bH_t \in \Cset^{N \times N_t}$, and $\bh_{r,k} \in \Cset^{N}$ represent the \gls{BS}-\gls{MU}, \gls{BS}-\gls{RIS}, and \gls{RIS}-\gls{MU} channel matrices, respectively. Moreover, $\bGamma \in \Cset^{N \times N}$ is a diagonal matrix whose $n$th diagonal entry is given by $[\bGamma]_n = [\bOmega]_n e^{\jj [\bomega]_n}$, where $[\bomega]_n$ and $[\bOmega]_n$ correspond to the phase shift and reflection amplitude applied by the $n$th \gls{RIS} unit cell. Throughout this paper, we assume $[\bOmega]_n=1,\,\forall n$.
\subsection{Channel Model}
\label{Channel Model}
Due to the assumption of an extremely large \gls{RIS}, the distances between the RIS and both the \gls{BS} and \glspl{MU} may fall within the \gls{NF} region of the RIS \cite{Liu2023nearfield,delbari2024nearfield}. Consequently, an \gls{NF} channel model is adopted. Additionally, since RISs are typically implemented at elevated positions, they maintain \gls{LOS} links with both the BS and \glspl{MU}. At high frequencies, these LOS links become even more dominant compared to \gls{nLOS} components. Therefore, the channels are modeled using Rician fading with a high $K$-factor, highlighting the strong contribution of \gls{LOS} components relative to \gls{nLOS} components. For generality, we define the channel model in terms of $\bH \in \Cset^{N_\rx \times N_\tx}$, where $N_\tx$ and $N_\rx$ denote the number of \gls{Tx} and \gls{Rx} antennas, respectively. A Rician \gls{MIMO} channel model is written as
\begin{equation}
\label{eq: channel model}
    \bH=\sqrt{\frac{K_f}{K_f+1}}\bH^{\mathrm{LOS}}+\sqrt{\frac{1}{K_f+1}}\bH^{\nLOS},
\end{equation}
where $K_f$, $\bH^{\mathrm{LOS}}$, and $\bH^{\nLOS}$ denote the $K$-factor, \gls{LOS}, and \gls{nLOS} \gls{NF} channels, respectively. $K_f$ determines the power ratio of the \gls{LOS} component to the \gls{nLOS} components of the channel. $\bH^{\mathrm{LOS}}$ and $\bH^{\nLOS}$ are given by:
\begin{IEEEeqnarray}{cc} 
	[\bH^\LOS]_{m,n} = \, c_0\e^{\jj\kk\|\bu_{\rx,m}-\bu_{\tx,n}\|}\label{Eq:LoSnear},\\
    \bH^\nLOS_s=c_s\ba_{\rx}(\bu_{s})\ba_{\tx}^\Trans(\bu_{s}),\\
 {[\ba_{\tx}(\bu_{s})]_n}   = \e^{\jj\kk\|\bu_{\tx,n}-\bu_{s}\|}
 \,\text{and}\, 
 {[\ba_{\rx}(\bu_{s})]_m}   = \e^{\jj\kk\|\bu_{\rx,m}-\bu_{s}\|}\!\!,\quad\label{Eq:nLoSnearPoint}
\end{IEEEeqnarray}
where $c_0$ denotes the amplitude of the \gls{LOS} path and $\kk=\frac{2\pi}{\lambda}$ is the wave number, where $\lambda$ is wavelength. The variables $\bu_{\tx,n}$ and $\bu_{\rx,m}$ correspond to the locations of the $n$th \gls{Tx} antenna and the $m$th \gls{Rx} antenna, respectively. The steering vectors $\ba_{\tx}(\cdot) \in \Cset^{N_{\tx}}$ and $\ba_{\rx}(\cdot) \in \Cset^{N_{\rx}}$ represent the \gls{Tx} and \gls{Rx} \gls{NF} array responses, respectively. Furthermore, $\bu_s$ indicates the location of the $s$th scatterer, and $c_s$ represents the end-to-end amplitude of the $s$th non-LOS path. Moreover, we assume direct link between \gls{BS} and \gls{MU} is blocked, i.e., $\bh_{d,k}\approx0,\,\forall k$.

\subsection{RIS Element Location Model}
\label{RIS Element Location Model}
In this study, we assume the center of the \gls{RIS} is the center of the global coordinate system. This \gls{RIS} lies on a non-planar surface extending along the $\y-\z$ axes. Each \gls{RIS} element's location can be obtained as $\bu_{\RIS}(g)=\bu^\rho_{\RIS}+[x,0,0]$ where $\bu^\rho_{\RIS}=[0,y,z], \,-\frac{L_y}{2}\leq y\leq\frac{L_y}{2},\,-\frac{L_z}{2}\leq z\leq\frac{L_z}{2}$ and $x=g(y,z)$ can be defined as a general function. For a planar array, we have $x=g(y,z)=0,\,\forall (y,z)$, whereas for an arbitrary \gls{RIS} geometry, $g(y,z)$ is unknown. The spatial placement of the \gls{RIS} elements is governed by a geometric function, denoted as $x = g(y,z)$. However, exact characterization of the \gls{RIS} surface geometry is generally an ill-posed problem without prior structural knowledge. To render the problem computationally tractable, we approximate the \gls{RIS} surface using its Taylor approximation:
\begin{equation}
    \label{eq: x wrt y and z Taylor}
    x=g(y,z)\approx \sum_{m_y=0}^{M_y}\sum_{m_z=0}^{M_\total-M_y} a_{m_y,m_z} y^{m_y}z^{m_z},
\end{equation}
where $0\leq M_y\leq M_\total$. This parametric approach significantly reduces the dimensionality of the estimation problem from $N$ (all element coordinates) to just the polynomial coefficients ($\frac{(M_\total+1)(M_\total+2)}{2}$). In this study, a second-order Taylor series approximation ($M_\total=2$) is sufficient to capture common structural curvatures such as cylindrical columns or parabolic facades, but \eqref{eq: x wrt y and z Taylor} can be used for any $M_\total$. By assuming $M_\total=2$, and omitting the constant part, we have a quadratic function of $y$ and $z$. This models the $x$-coordinate (depth) of the surface as:
\begin{equation}
    \label{eq: x wrt y and z quadratic}
    x=g(y,z)\approx a_{yy}y^2+a_{zz}z^2+a_{yz}y z+a_{y}y+a_{z}z,
\end{equation}
where the coefficients $a_{yy}$, $a_{zz}$, $a_{yz}$, $a_{y}$, and $a_{z}$ are fixed but unknown parameters for a given deployment that must be estimated. While the exact function $g$ is unknown in practical scenarios, the physical dimensions (ranges of $y$ and $z$) are typically known. In the subsequent section, we demonstrate how this geometric model is utilized to configure the \gls{RIS} phase shifts to satisfy a quality of service requirement for the \glspl{MU}.
\section{Problem formulation and Solution}
\label{sec: Machine Learning Approach}
To optimally configure the \gls{RIS} phase shifts, accurate estimation of the surface geometry parameters defined in \eqref{eq: x wrt y and z quadratic} is a prerequisite. We adopt a data-driven approach, utilizing a sampling dataset composed of \gls{MU} locations and the corresponding \gls{RIS} configuration phase shifts. We rely exclusively on the received power measurements rather than complex channel state information. This design choice is motivated by the fact that phase measurements are often unstable and require complex synchronization hardware, whereas power measurements are robust and easily accessible \cite{delbari2022noncoherent}. The noiseless received power via the \gls{LOS} path is\footnote{In \gls{mmWave}, the impact of \gls{nLOS} paths relative to the \gls{LOS} path is negligible, especially when narrow beams are adopted.}
\begin{equation}
    \label{eq: power}
    P(k,\bomega;g)\!=\!\Big|\alpha_0\int_{-\frac{L_y}{2}}^{\frac{L_y}{2}}\int_{-\frac{L_z}{2}}^{\frac{L_z}{2}}\e^{\jj\omega(y,z)+\jj \kappa d_{k}(y,z;g)}\dd y\dd z\Big|^2,
\end{equation}
where $d_{k}(y,z;g)=\kappa\|\bu_{\RIS}(g)-\bu_\BS\|+\kappa\|\bu_{\RIS}(g)-\bu_{\MU,k}\|$, and $\alpha_0$ captures the total pathloss. The data collection involves $K$ distinct \gls{MU} locations. For each location $k$, the \gls{RIS} cycles through $M$ different phase configurations, denoted as $\bomega_m$ for $m=1,\dots, M$. The \gls{BS} sends a signal and \gls{MU} measures the resulting signal power for each combination, $P_{k,m} = |y_{k}(\bomega_m)|^2$. The structure of the simulated collected training dataset is summarized in Table~\ref{tab: phase and location}.
\begin{table}[htbp]
\centering 
\begin{tabular}{| c | c  c  c  c|}
\hline
\diagbox{$k$}{$\bomega$} & \textbf{$\bomega_1$} & \textbf{$\bomega_2$} & \textbf{$\dots$} & \textbf{$\bomega_M$} \\
\hline
1 & $P_{1,1}$ & $P_{1,2}$ & $\dots$ & $P_{1,M}$ \\
2 & $P_{2,1}$ & $P_{2,2}$ & $\dots$ & $P_{2,M}$ \\

$\vdots$ & $\vdots$ & $\vdots$ & $\ddots$ & $\vdots$ \\

$K$ & $P_{K,1}$ & $P_{K,2}$ & $\dots$ & $P_{K,M}$ \\
\hline
\end{tabular}
\caption{Received power measurements ($P_{k,m}$) across $K$ user locations and $M$ different RIS phase configurations.}
\label{tab: phase and location}
\end{table}
The \gls{NLS} problem is formulated as:
\begin{align}
    \hat{g}=&\underset{g}{\arg\min} \quad \sum_{k=1}^K\sum_{m=1}^M\Big\|P_{k,m}-P(k,\bomega;g)\Big\|^2.
\end{align}
Solving this optimization problem by \gls{NLS} directly for every \gls{RIS} deployment is computationally intensive. To address this, we introduce an \gls{NN} that incurs a one-time training cost but allows for rapid inference during implementation. Furthermore, the \gls{NN} architecture is generalized, maintaining a consistent structure even if the underlying function changes. The dataset in Table~\ref{tab: phase and location} is utilized to train an \gls{NN} regressor. The \gls{NN} takes the received power vector and user coordinates as input and estimates the five geometric coefficients $\{a_{yy}, a_{zz}, a_{yz}, a_{y}, a_{z}\}$. Once these coefficients are estimated, the \gls{RIS} surface is effectively "known," allowing for the analytical or numerical optimization of phase shifts for any future user location.

\subsection{Data Collection and Measurement Point Spacing}
\label{sec:MeasurementPoints}
\begin{figure*}[t]
\begin{subfigure}{0.3\textwidth}
    \centering
    \includegraphics[width=1\textwidth]{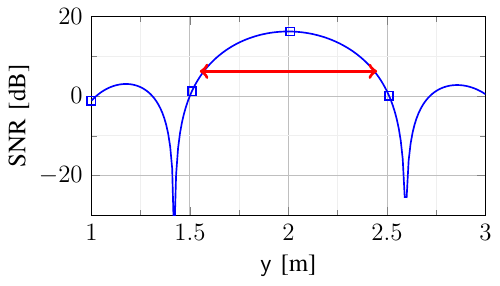}
    \caption{The received power along to the $\y$ axis.}
    \label{fig: snr y}
\end{subfigure}
\hfill
\begin{subfigure}{0.3\textwidth}
    \centering
    \includegraphics[width=1\textwidth]{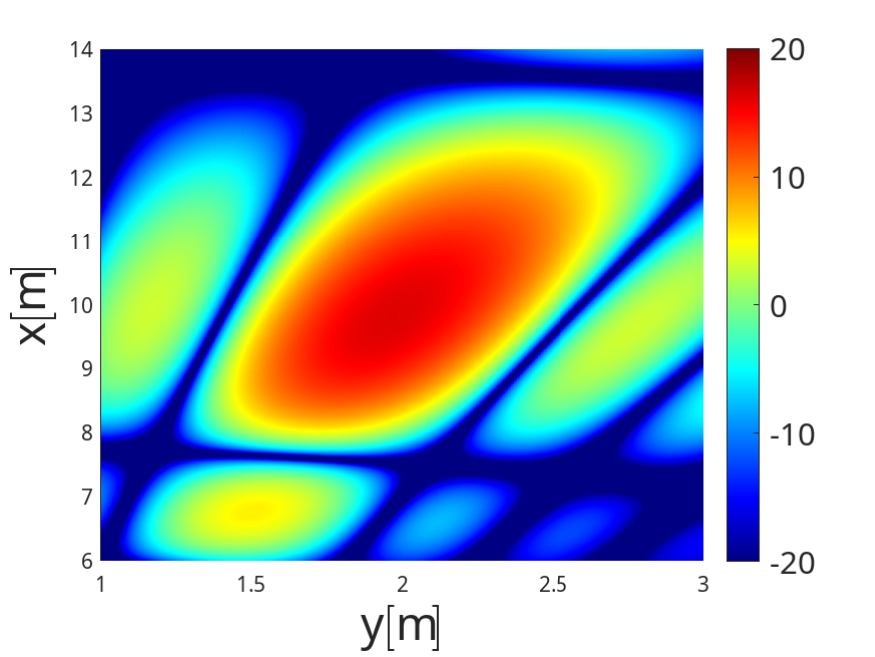}
    \caption{The received power inside the coverage area.}
    \label{fig: measurement}
\end{subfigure}
\hfill
\begin{subfigure}{0.3\textwidth}
    \centering
    \includegraphics[width=1\textwidth]{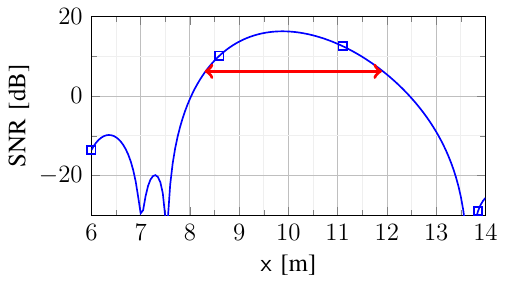}
    \caption{The received power along to the $\x$ axis.}
    \label{fig: snr x}
\end{subfigure}
\caption{The required minimum distance for the spatial density of the measurements.}
\label{fig: spatial density}
\vspace{-5mm}
\end{figure*}

The spatial density of measurement points significantly impacts estimation accuracy. To ensure the measurements provide distinct spatial information, the sampling points must be separated sufficiently. For a large array, the normalized beam pattern can be approximated by a sinc function, $\sinc(x) = \sin(\pi x)/(\pi x)$, whose envelope decays proportional to $1/|x|$. To ensure the received power at a neighboring sample point drops sufficiently, e.g., 10 dB relative to the peak, we utilize the decay property of the sinc envelope. A standard heuristic for this distance is approximated by $d_t \approx \frac{2r}{N_t}$, where $r$ is the distance of the received power location from the \gls{RIS} and $N_t$ is the number of the \gls{RIS} elements in axis $t,\,\forall t=\{y,z\}$. For our simulation setup, we have $N_y=40$, $N_z=10$, and a distance of $r=10~$m. These parameters lead to distances of $\frac{2\times10}{40}=0.5~$m and $\frac{2\times10}{10}=2$~m in $\y$ and $\z$ directions, respectively. These values are confirmed by Fig. \ref{fig: spatial density}.

\subsection{Network Architecture and Training}
\label{sec: Network Architecture and Training}

The data set comprises $S$ independent samples generated via the measurement procedure described in Section~\ref{sec: Machine Learning Approach}. Each sample pairs an input vector of received powers, $\mathbf{x} \in \mathbb{R}^{K}$, derived in \eqref{eq: power}, with a target vector of geometric coefficients, $\mathbf{y} \in \mathbb{R}^{5}$.

\subsubsection{Preprocessing and Data Partitioning}
To ensure numerical stability and efficient convergence, the input power measurements are normalized to the range $[0, 1]$ via min-max scaling. 
The processed dataset is partitioned into training (80\%), validation (10\%), and testing (10\%) sets.

\subsubsection{Architecture and Hyperparameters}
We employ a feedforward \gls{MLP} to model the non-linear mapping between received power and surface geometry. The network consists of an input layer of dimension $K$, followed by three fully connected hidden layers containing 128, 64, and 32 neurons, respectively. This tapering structure acts to progressively compress the features relevant for geometry extraction.

\Gls{ReLU} activation is applied to all hidden layers, while the output layer utilizes a linear activation function to regress the five continuous parameters of the quadratic surface model discussed in \eqref{eq: x wrt y and z quadratic}. The network is trained using the Adam optimizer to minimize the \gls{MSE} loss. Training is set for a maximum of 500 epochs with a mini-batch size of 100. To prevent overfitting, an early stopping mechanism terminates training if the validation loss fails to improve for 50 consecutive epochs.

\subsection{RIS Phase Shift Design}
\label{sec: RIS phase shift design}
While various techniques exist to optimize \gls{RIS} phase shifts, $\bomega_1,\cdots,\bomega_M$, our choice is to optimize them such that if \gls{RIS} would be planar, then all reflected signals add up coherently in the targeted location. Let us assume \gls{RIS} phase shifts must be optimized for the $k'$th location, then
\begin{equation}
\label{eq: phase shift design}
    \omega(y,z)=-\kappa\|\bu^\rho_{\RIS}-\bu_\BS\|-\kappa\|\bu^\rho_{\RIS}-\bu_{\MU,k'}\|.
\end{equation}
We substitute \eqref{eq: phase shift design} into \eqref{eq: power} and make our data set for the \gls{NN}.
\section{Performance Evaluation} \label{Performance Evaluation}
\subsection{Simulation Setup}
We employ the simulation setup illustrated in Fig. \ref{fig:system model}. The \gls{RIS} is centered at the origin of the Cartesian coordinate system, i.e., [0,0,0]~m. We assume $K$ different measurement locations inside a fixed area in $\Pset_u\in\{(\x,\y,\z):6~\text{m}\leq\x\leq 14~\text{m}, 1~\text{m}\leq\y\leq 3~\text{m}, \z=-5\}$. The \gls{BS} comprises a $4\times4=16$ \gls{UPA} positioned along the $\x-\z$ plane, and located at $[40,20,5]~\text{m}$. The \gls{RIS} is a uniform non-planar array consisting of $N_\y\times N_\z=40\times10$ elements aligned to the $\y$ and $\z$ axes, respectively. The element spacing for both the \gls{BS} and \gls{RIS} is half of the wavelength. The noise variance is computed as $\sigma_n^2=WN_0N_{\rm f}$ with $N_0=-174$~dBm/Hz, $W=20$~MHz, and $N_{\rm f}=6$~dB. We assume $28$~GHz carrier frequency, and $\rho(d_0/d)^\sigma$ pathloss model where $\rho=-61$~dB at $d_0=1$~m. Moreover, we adopt $S=18,225$, $K=25$, $\sigma = (2,2,2)$ and  $K_f=(0,10,10)$ for the \gls{BS}-\gls{MU}, \gls{BS}-\gls{RIS}, and \gls{RIS}-\gls{MU} channels, respectively. The training parameters are given in Table~\ref{tab:hyperparameters}. Performance was quantified using \gls{MSE} which is a standard regression metric. This metric was calculated independently for each of the five output parameters to assess the model's predictive accuracy on a per-parameter basis.

\begin{table}[t]
\centering
\caption{Neural Network Hyperparameters}
\label{tab:hyperparameters}
\begin{tabular}{ll}
\toprule
\textbf{Hyperparameter} & \textbf{Value} \\
\midrule
\multicolumn{2}{l}{\textit{Data Partitioning}} \\
Training Set Size     & 80\% \\
Validation Set Size   & 10\% \\
Test Set Size         & 10\% \\
\midrule
\multicolumn{2}{l}{\textit{Network Architecture}} \\
Hidden Layers         & 3 (Dense) \\
Neuron Configuration  & 128 $\rightarrow$ 64 $\rightarrow$ 32 \\
Hidden Activation & ReLU \\
Output Activation & Linear \\
\midrule
\multicolumn{2}{l}{\textit{Training Configuration}} \\
Optimizer             & Adam \\
Learning Rate         & 0.001 \\
Loss Function         & MSE \\
Maximum Epochs        & 500 \\
Batch Size            & 100 \\
Early Stopping Patience & 50 Epochs \\
\bottomrule
\end{tabular}
\end{table}

\subsection{Simulation Results}
\begin{figure}
    \centering
    \includegraphics[width=0.4\textwidth]{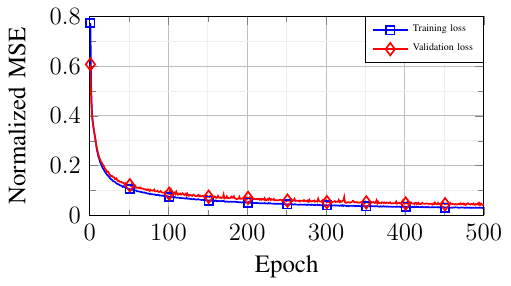}
    \caption{Training loss and validation loss versus the number of epochs.}
    \label{fig:training}
    \vspace{-0.5 cm}
\end{figure}

First, we study the convergence behavior of the training algorithm in Fig. \ref{fig:training}. We plot normalized \gls{MSE} versus the number of epochs. As the number of epochs increases, the normalized \gls{MSE} of both the training loss and the validation loss decreases smoothly, confirming convergence.

\begin{figure}
    \centering
    \includegraphics[width=0.4\textwidth]{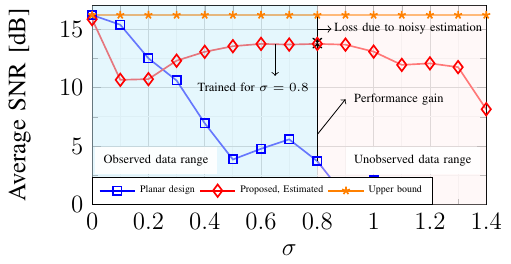}
    \caption{\gls{Rx} power at the \gls{MU} versus the geometric parameter variance.}
    \label{fig:power_sigma}
    \vspace{-0.4 cm}
\end{figure}

Fig.~\ref{fig:power_sigma} shows the \gls{Rx} \gls{SNR} (dB) for different $\sigma$ values under three different designs. Here, $\sigma$ is a parameter defining the upper bound of the uniform distribution from which $a_{yy}$, $a_{yz}$, and $a_{zz}$ are drawn, i.e., $a_{t_1t_2}\sim \Uset(0,\sigma),\,\forall t_1, t_2 \in\{y,z\}$. Since the \gls{RIS} geometry is highly sensitive to linear terms ($a_y, a_z$), these parameters were drawn from a narrower distribution, $\Uset(-\frac{\sigma}{10},\frac{\sigma}{10})$. The training data set is optimized for $\sigma=0.8$. The blue curve represents the conventional \gls{RIS} planar model assumption, while the orange curve serves as an upper bound, assuming perfect knowledge of the element locations. Finally, the red curve corresponds to the proposed algorithm. All three methods achieve peak performance at $\sigma=0$, where the \gls{RIS} geometry is purely deterministic and planar. In the low-variance regime ($0 < \sigma \leq 0.25$), the planar design remains sufficient and slightly outperforms the proposed algorithm, as the latter is specifically optimized for $\sigma=0.8$. However, for $\sigma \geq 0.25$, the planar model's performance degrades significantly compared to the proposed design. Notably, the algorithm demonstrates robustness by degrading slowly as $\sigma$ enters the unobserved data region ($\sigma > 0.8$).

\begin{figure}
    \centering
    \includegraphics[width=0.4\textwidth]{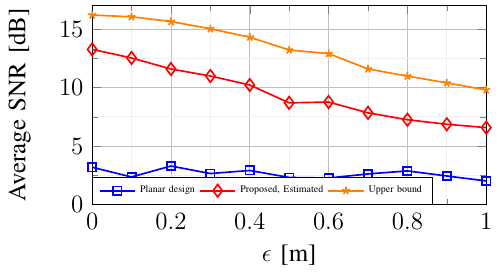}
    \caption{The \gls{Rx} power in \gls{MU} versus the variance of the localization error.}
    \label{fig:SNR_error}
    \vspace{-0.5 cm}
\end{figure}

Fig.~\ref{fig:SNR_error} illustrates the average \gls{SNR} (dB) at the \gls{MU} in the presence of location estimation errors. The estimated location of the \gls{MU} is modeled as $[\x+\epsilon_x,\y+\epsilon_y,-5]$~m, where both $\epsilon_x$ and $\epsilon_y$ are drawn from a uniform distribution $\Uset(-\epsilon,\epsilon)$. As observed, the average \gls{SNR} decreases as the error increases for both the proposed and the upper bound. Interestingly, the planar design is not significantly affected by location error. This is because, due to the model mismatch, the \gls{RIS} beam does not point towards the \gls{MU} in the first place, and further location errors do not significantly affect the received power. The proposed scheme significantly outperforms the planar design even under certain location errors. Utilizing more robust \gls{RIS} phase-shift designs to mitigate these errors \cite{delbari2024far} is reserved for future work.

\section{Conclusion} \label{s:Conclusion}
In this paper, we investigated the problem of beamforming optimization for \glspl{RIS} deployed on non-planar surfaces, a practical scenario where conventional planar assumptions lead to performance degradation. To address the limitations of existing analytical models, which are often restricted to specific geometries, we proposed a robust \gls{DL}-based framework. We first developed a parametric model to efficiently represent surface curvature with a reduced number of variables. Subsequently, we designed an \gls{NN} capable of implicitly estimating the surface geometry and optimizing phase shifts using only sparse \gls{Rx} power measurements. Simulation results confirmed that the proposed approach not only converges efficiently but also yields performance gains over planar beamforming designs, particularly as the variations in element location increase.

\bibliographystyle{IEEEtran}
\bibliography{References}

\end{document}

%% file: preamble.tex
\usepackage{amsfonts,amsmath,amsthm,bm}
\usepackage{cite}
\usepackage{amsmath,amssymb,amsfonts}
\usepackage{algorithmic}
\usepackage{graphicx}
\usepackage{textcomp}
\usepackage{xcolor}
\usepackage[T1]{fontenc}
\usepackage{comment}
\usepackage{eucal}
\usepackage{algorithm}
\usepackage{algorithmic}
\usepackage[acronyms,nonumberlist,nopostdot,nomain,nogroupskip]{glossaries}

\usepackage{subcaption}
\usepackage[font=footnotesize]{caption}
\usepackage{booktabs}
\usepackage{hyperref}
\hypersetup{
    colorlinks=true,        
    linkcolor=blue,         
    citecolor=blue,          
    urlcolor=black
}
\usepackage{orcidlink}
\usepackage{diagbox}  

\newtheorem{lem}{Lemma}

\def\bOmega{\boldsymbol{\Omega}}
\def\bmu{\boldsymbol{\mu}}
\newcommand{\e}{\mathsf{e}}
\newcommand{\jj}{\mathsf{j}}
\newcommand{\Herm}{\mathsf{H}}
\newcommand{\Trans}{\mathsf{T}}
\newcommand{\x}{\mathsf{x}}
\newcommand{\y}{\mathsf{y}}
\newcommand{\z}{\mathsf{z}}
\newcommand{\bx}{\mathbf{x}}

\newcommand{\bH}{\mathbf{H}}
\newcommand{\ba}{\mathbf{a}}
\newcommand{\bA}{\mathbf{A}}

\newcommand{\bu}{\mathbf{u}}

\newcommand{\bh}{\mathbf{h}}

\newcommand{\bq}{\mathbf{q}}

\newcommand{\kk}{\kappa}

\newcommand{\bSigma}{\boldsymbol{\Sigma}}
\newcommand{\bGamma}{\boldsymbol{\Gamma}}

\newcommand{\Ex}{\mathbb{E}}

\newcommand{\dd}{\mathrm{d}}

\newcommand{\tx}{\mathrm{tx}}

\newcommand{\total}{\mathrm{total}}

\def\bomega{\boldsymbol{\omega}}

\def\Cset{\mathbb{C}}

\def\Rset{\mathbb{R}}

\def\LOS{\mathrm{LOS}}
\def\nLOS{\mathrm{NLOS}}

\def\tx{\mathrm{tx}}
\def\rx{\mathrm{rx}}
\def\BS{\mathrm{BS}}
\def\MU{\mathrm{MU}}
\def\RIS{\mathrm{RIS}}

\def\sinc{\mathrm{sinc}}

\def\sCN{\mathcal{CN}}

\def\Pset{\mathcal{P}}

\def\Uset{\mathcal{U}}

\usepackage{xcolor}

\usepackage{microtype}

\input{acronyms}

%% file: acronyms.tex
\newacronym{RIS}{RIS}{reconfigurable intelligent surface}
\newacronym{QoS}{QoS}{quality of service}
\newacronym{SNR}{SNR}{signal to noise ratio}
\newacronym{BS}{BS}{base station}
\newacronym{MU}{MU}{mobile user}
\newacronym{NF}{NF}{near-field}
\newacronym{FF}{FF}{far-field}
\newacronym{Tx}{Tx}{transmitter}
\newacronym{Rx}{Rx}{receiver}
\newacronym{AWGN}{AWGN}{additive white Gaussian noise}
\newacronym{wrt}{w.r.t.}{with respect to}
\newacronym{UPA}{UPA}{uniform planar array}
\newacronym{RF}{RF}{radio frequency}
\newacronym{LOS}{LOS}{line of sight}
\newacronym{nLOS}{NLOS}{non-LOS}
\newacronym{MISO}{MISO}{multi-input single-output}
\newacronym{MIMO}{MIMO}{multiple-input multiple-output}
\newacronym{CSI}{CSI}{channel state information}
\newacronym{NN}{NN}{neural network}
\newacronym{DL}{DL}{deep learning}
\newacronym{CNN}{CNN}{convolutional neural network}
\newacronym{MSE}{MSE}{mean squared error}
\newacronym{ULA}{ULA}{uniform linear array}
\newacronym{mmWave}{mmWave}{millimeter-wave}
\newacronym{MLP}{MLP}{multi-layer perceptron}
\newacronym{ReLU}{ReLU}{rectified linear unit}
\newacronym{LS}{LS}{least square}
\newacronym{NLS}{NLS}{non-linear least squares}